# Shapeshifting diffractive optical devices

**Authors:** S.L. Oscurato[1,2,3,*], F. Reda[1], M. Salvatore[1], F. Borbone[4,2], P. Maddalena[1,2,3], A. Ambrosio[2,*].

## Affiliations

[1] Physics Department "E. Pancini", University of Naples "Federico II", Complesso Universitario di Monte Sant'Angelo, Via Cintia 21, 80126, Naples, Italy.

[2] CNST@POLIMI—Fondazione Istituto Italiano di Tecnologia, Via Pascoli 70, 20133 Milan, Italy.

[3] Centro Servizi Metrologici e tecnologici Avanzati (CeSMA), University of Naples "Federico II", Complesso Universitario di Monte Sant'Angelo, Via Cintia 21, 80126, Naples, Italy.

[4] Department of Chemical Sciences, University of Naples "Federico II", Complesso Universitario di Monte Sant'Angelo, Via Cintia 21, 80126, Naples, Italy.

*Correspondence to: stefanoluigi.oscurato@unina.it; antonio.ambrosio@iit.it;

## Abstract

In optical devices like diffraction gratings and Fresnel lenses, light wavefront is engineered through the structuring of device surface morphology, within thicknesses comparable to the light wavelength. Fabrication of such diffractive optical elements involves highly accurate multi-step lithographic processes that in fact set into stone both the device morphology and optical functionality. In this work, we introduce shapeshifting diffractive optical elements directly written on an erasable photoresist. We first develop a lithographic configuration that allows writing/erasing cycles of aligned optical elements directly in the light path. Then, we show the realization of complex diffractive gratings with arbitrary combinations of grating vectors. Finally, we demonstrate a shapeshifting diffractive lens that is reconfigured in the light-path in order to change the imaging parameters of an optical system.

# Introduction

Diffraction gratings are among the first optical devices ever realized. Light structuring through gratings results from the periodic phase modulation accumulated across the grating profile (*1*). A diffraction grating made of periodically spaced grooves produces, far from the grating, a line of light dots (diffraction orders) along a direction perpendicular to the grating grooves (grating vector). Although simple, diffraction gratings represent the first example of structuring light by means of an engineered phase modulation. This concept is at the core of Holography and Fourier optics (when the field structuring can be considered as only due to the propagation of component plane waves) (*2*). Moreover, phase masks can be multiplexed, so gratings with different Fourier components can be combined together to achieve more complex bi-dimensional light structuring (*3*).

Diffraction gratings represent also the first example of planar optics. For a grating with sinusoidal profile made of a 1.7 refractive index transparent material, the maximum diffraction efficiency is achieved with a morphology modulation of 0.84 times the illuminating light wavelength (i.e. only 531 nm thickness for an illuminating light of 633 nm) (*4*). Other diffractive optical elements (DOEs) of common use are Fresnel lenses, in practice, cylindrically symmetric gratings optimized to focus collimated light at a design distance (*5*). Such lenses, used for instance in lighthouses, become handy to reduce space and weight constraints related to standard refractive lenses (*6, 7*). Once the optical parameters of the material are fixed, the optical functionality of a DOE is controlled through the device morphology, in case of a grating: spacing, height and profile of the grooves. Such design principle has been continuously developed for DOEs manufacturing, resulting in optimized elements for specific applications. An interesting discussion about advanced diffraction elements for spectroscopy is provided by the recent ref. (*8*). However, most of the DOEs manufacturing techniques consist of at least two steps, typically a molding process followed by a development stage (and coating in case of reflective DOEs) (*4*). All the different realization approaches also result in static DOE, with fixed geometry and optical functionality. Here we report the realization of diffractive optical elements whose morphology can be changed in real time to provide different on-demand optical functionalities. Phase and amplitude masks for diffractive optics can be realized with different approaches. On one hand, there are fast but pixeled modulators, spanning from commercially available liquid-crystal-based Spatial Light Modulators (SLM) or Digital Mirror Devices (DMDs) (*9*) to recent 2D materials devices (*10*). On the other hand, in the last ten years, research on optical metasurfaces exploded, providing new devices with unprecedented modulation possibilities and resolution (*11–13*) but, in fact, much limited in terms of optical tunability (*14*).

In this work we develop a holographic photo-lithography scheme employing a reconfigurable polymeric photoresist to realize diffractive optical elements that can be re-shaped directly in the light-path to provide a specific optical functionality; in fact, shapeshifting diffractive optical elements that do not need any additional development step. Complex gratings are showed together with a reconfigurable monochromator. Finally, variable focal lengths diffractive lenses are demonstrated that allow magnification tuning of a complex optical system.

**Results and Discussion**

The central equation to design a phase mask $\varphi(x, y)$ for a transmissive DOE is:

$$\varphi(x, y) = k_0 (n - n_0) h(x, y) \qquad (1)$$

Once the refractive indices of the grating material ($n$) and the surrounding material ($n_0$) are fixed, together with the illuminating light wavelength $\lambda_0$ (wavevector $k_0 = 2\pi/\lambda_0$), the phase mask is entirely defined by the surface morphology $h(x, y)$.

Our photoresist is a polymer containing azobenzene molecules (*15*, *16*) (see details in Materials and Methods in supplementary materials). When illuminated by a specific light pattern, the polymer surface is deformed resulting in a morphology that follows the illuminating light distribution (*17–19*). Illuminating such polymer with a 2D light pattern with sinusoidal profile, results in a surface with similar profile, i.e. a diffraction grating (see also Fig. S3). The process does not require any further development step. In this case, the periodic light distribution can be as simple as that produced by two interfering light beams (*20–23*). In our experiment instead, we used a holographic setup that projects complex and arbitrary light patterns on the polymer surface (*24*). Such patterns can also be changed in time (details of our experimental setup can be found in supplementary materials).

Figure 1 shows our writing process. The writing beam is a holographically patterned laser beam with a wavelength of 491 nm (circularly polarized). The generated light pattern is projected onto the polymer film surface by means of a 50X long working distance objective (Fig. 1A). The appearance of a diffraction grating on the surface can be monitored in real time by detecting the light power in the first diffraction order of a probe beam at wavelength of 633 nm, collimated through the same objective. Figure 1E shows the rising diffraction efficiency during the writing process. Another circularly polarized laser beam with wavelength of 405 nm) also illuminates the grating area, from the substrate side. The function of the 405 nm beam is twofold (see also supplementary materials): at low power density (~0.5 $W/cm^2$) it keeps *active* the photo-sensitive molecules in the polymer, avoiding partial saturation, assisting and speeding-up the writing process (*25*); at high power density (~1 $W/cm^2$), the violet beam erases the polymer surface (*26*, *27*) (see also Fig. S4), that is then ready to write the next DOE. It is important to note that, once the writing process is concluded, the realized DOEs are still stable for years at room temperature and ambient illumination conditions. Figure 1F and H show the Atomic Force Microscope (AFM) micrograph of the polymer film surface after writing and after erasing, respectively. Figure 1E also shows the decreasing of the diffraction efficiency over the erasing process: both the morphology and diffraction residuals are negligible. At this point, a new DOE can be realized on the film surface; for instance, a new grating with different pitch (Fig. 1J), to be later eventually erased again (Fig. 1L). In our experimental conditions, we noticed no significant degradation of the surface for at least 15 writing-erasing cycles (see also Fig. S6).

As next step, we tested the possibility of having a reconfigurable DOE in the light-path to realize a reconfigurable monochromator. In this case, the light diffracted by the grating is detected by means of a CCD camera. The probe beam is a white light beam provided by a LED. Figure 1N-R show the effect of re-shaping the grating periodicity. As expected, the smaller the grating pitch, the higher the diffractive power and the separation between the light spectral components. In this case, changing the DOE in the light-path allows real time tuning of the spectrometer resolution through the grating dispersion. In the same configuration, fixing an aperture on the camera (same as using a physical aperture and a point detector) allows to observe how the wavelength diffracted in the detection area changes as a function of the realized grating. This last configuration works in fact as a tunable monochromator, without any moving part, that can be used together with a broadband source to select an illuminating wavelength band while preserving the original alignment.

Another unique characteristic of our lithographic approach stands in the noise reduction of the holographic light pattern and the resulting enhanced quality of the morphology on the photoresist surface. Usually, light structuring through digital holography suffers of a random speckles distribution that degrades the contrast and limits the achievable gray levels in the projected light pattern, also compromising the definition of small features. To minimize such effects, we continuously refresh the light pattern at a refresh rate of 20 Hz. Each new light pattern has a random speckles distribution that averages down during the writing time, improving in fact the quality of the structured polymer morphology (*24*). For instance, in order to realize a grating with a single vector (like that of Fig. 1F-J), we expose the polymer to the required light pattern with random speckles distribution changing every 50 ms over the writing time. The writing beam power can be set to values that make the photoresist structuring dynamics slow enough to respond in fact to an averaged light pattern.

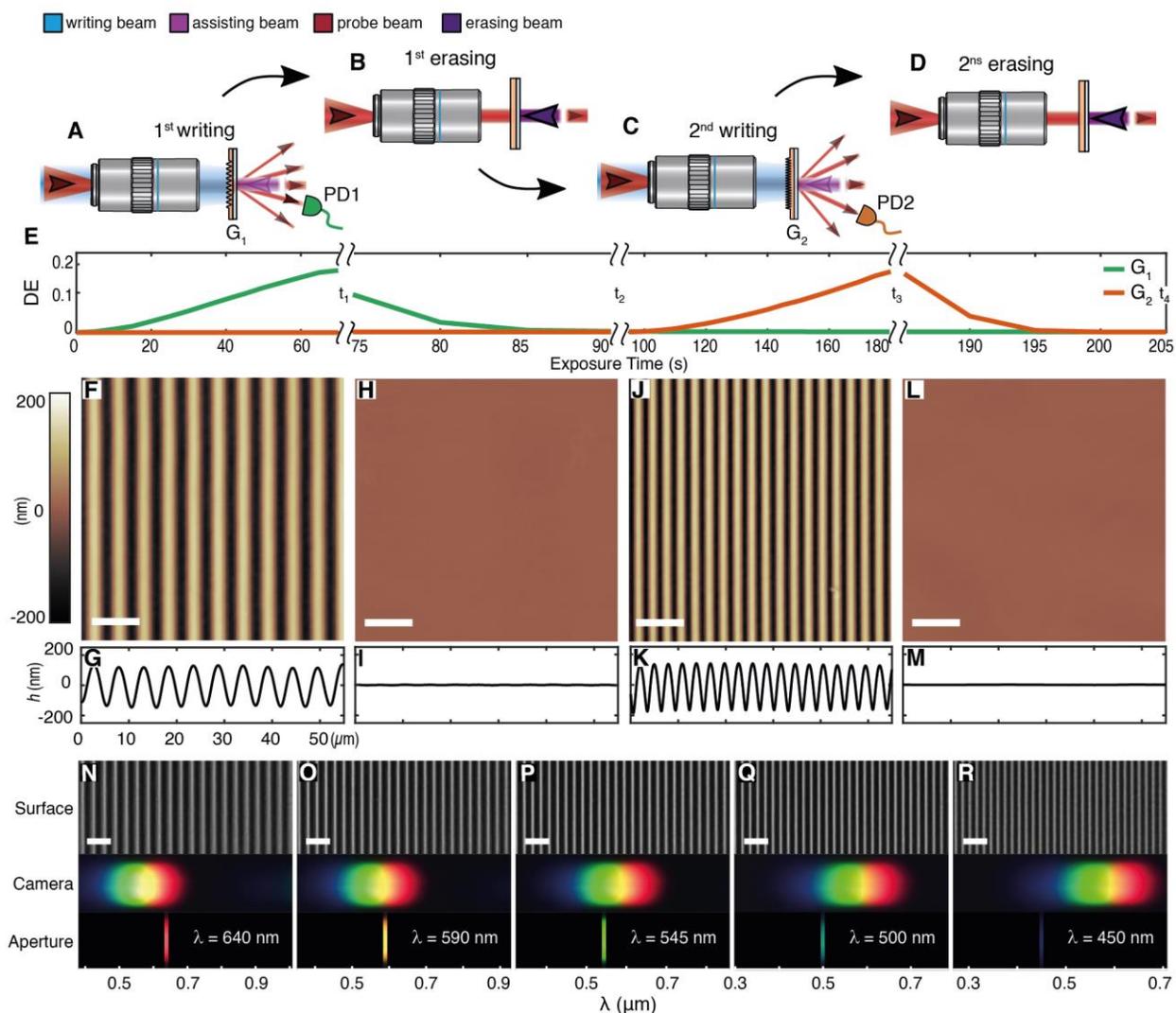

**Fig. 1. Direct all-optical realization of reconfigurable diffraction gratings.** (**A** to **D**) Schematic illustrations of optical configurations used for the writing, erasing, and the real-time monitoring of dynamically reconfigurable diffraction gratings. In the writing steps (A and C), the writing beam (light blue color) illuminates a photoresist area ($\sim 200 \, \mu m$ in diameter) with holographically-controlled sinusoidal intensity patterns through a 50X (NA=0.55) microscope objective. A collimated beam ($\lambda = 405 \, nm$) is used at low intensity to accelerate the surface structuration (light violet beam) and at higher intensity (dark violet beam) in the erasing steps (B and D), when the writing beam is switched off. (**E**) Time-evolving first-order diffraction efficiency curves recorded by two photodiodes (PD1 and PD2 shown in panels A and C) for a He-Ne probe beam (red beam) during the reconfiguration of two gratings of different periodicity (green curve for $G_1$ and orange curve for $G_2$). (**F** to **M**) AFM micrographs (F-H-J-L) and relative horizontal topographic profiles (G-I-L-M) of the surface at the instants ($t_i$) of the time sequence in (E). Scale bars in AFM images are $10 \, \mu m$. (**N** to **R**) Dynamical tuning of diffraction dispersion obtained through multiple grating periodicity reconfigurations. Top panels show optical micrographs of the surface (scale bar $10 \, \mu m$). The colored diffraction patterns produced by the surface from a white LED source are imaged by a CCD camera in the surface Fourier conjugate plane. A fixed aperture on the CCD camera intercepts color bands of shifting central wavelength.

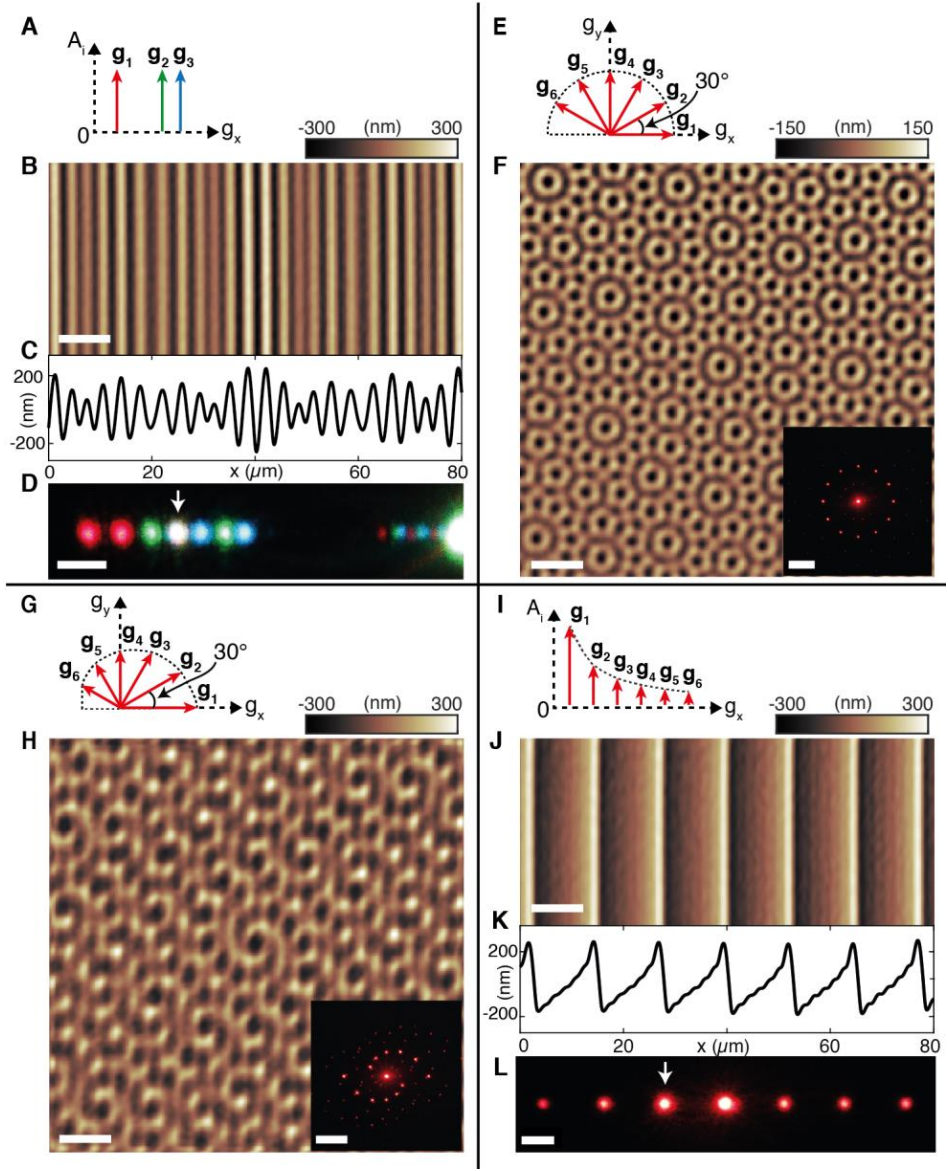

**Fig. 2. Diffraction gratings with arbitrary combination of grating vectors.** Experimental details and design parameters can be found in Table S1 in supplementary materials. (**A** to **D**) Three-component RGB diffraction grating. (A) The surface is designed as superposition of three sinusoids of periodicity $\Lambda_{x,i}$ (grating vectors $g_{x,i} = 2\pi/\Lambda_{x,i}$) and equal amplitude $A_i$. Measured AFM micrograph (B) and topographic profile (C) of the surface. (D) Photograph of the far-field diffraction pattern produced by the surface upon simultaneous illumination with three collinear laser beams at design wavelengths. The white arrow indicates the RGB diffraction order. (**E** to **H**) Two-dimensional diffraction gratings obtained as superposition of six sinusoidal functions. (F and H) AFM micrographs of quasicrystal surfaces designed with grating vector distributions schematized in (E and G), respectively. Photographs of far-field diffraction patterns, produced (on a screen at distance $d = 8.5\ cm$) when the diffractive elements are illuminated with a He-Ne laser beam, are shown as inset (scalebar $1.0\ cm$). (**I** to **L**) Blazed grating realized as superposition of six grating vectors of length $g_{x,i} = ig_1$ and amplitudes $A_i \sim A_1/i$ (I). (J) AFM and (K) topography of the surface. The grating is designed to direct most of the diffracted light in one of the first diffraction orders (highlighted by an arrow in the diffraction photograph in (L)). The power in the design diffraction order is more than 6 times higher than in the other first order. Scale bars in AFM micrographs are $10\ \mu m$; scale bars in D and L are $0.2\ cm$.

An identical approach can be used to realize DOEs with any distribution of grating vectors, directly encoded in the analytical design of grayscale digital holograms (see also Fig. S7). Figure 2 shows some examples. First a RGB grating, designed as even superposition of three grating vectors (Fig. 2A), is realized (Fig. 2B) to diffract three different wavelengths ($\lambda_1 = 633\ nm$, $\lambda_2 = 532\ nm$ and $\lambda_3 = 488\ nm$) into the same diffraction order (Fig. 2D).

Figure 2F shows a two-dimensional quasicrystal structure (*28*) characterized by 6 grating vectors $g_i$, differently oriented in the transverse plane (Fig. 2E). Figure 2H shows the effect of having same vectors orientation than Figure 2F but different vectors lengths (different periodicity) in the designed superposition of sinusoidal functions. This results into a spiral quasicrystal DOE.

Figure 2I shows the realization of a blazed grating that can diffract a probe beam into a preferential diffraction order. In this case, the blazed structure results from combining 6 gratings vectors with same direction but different lengths and weighted amplitudes. Besides being reconfigurable, DOE structures like those showed in Figure 2 are comparable with the state of the art of the most recent multiplexed structures obtained by static lithography techniques (*29*).

Related to the quality of the obtained structures, it is to be considered that the polymer that we used is fully compatible with a further lithographical step, where the obtained morphology can be transferred to a PDMS stamp for high quality replicas (see also Fig. S7) that provides our process with technological significance as new photolithography technique to be applied to optical systems as well as functionalized surfaces and microfluidics.

To further highlight the potential of our approach, in Figure 3 and 4 we show the realization of reconfigurable diffraction lenses working in real imaging systems. DOEs with such profiles are known as Gabor phase zone plates (*30*). Figure 3A shows the computed profile for a lens with $500\ \mu m$ focal length and numerical aperture $NA = 0.19$. The light distribution in the *yz* plane around the focal position can be calculated through the Fresnel diffraction integral (*2*). The calculated diffraction limited spot of this lens has FWHM of $\approx 1.7\ \mu m$ when illuminated by a beam with 633 nm wavelength (Fig. 3F). The simulated features are well reproduced in the measured intensity distribution (Fig. 3H and I), obtained from the experimental diffractive lens shown in Fig. 3B. Figure 3J also compares the predicted and realized efficiency for this DOE as a function of average grooves height, achieved in different exposure times. For this planar lens, the predicted maximum efficiency of ~32% is achieved in 50 s exposure (Fig. S8).

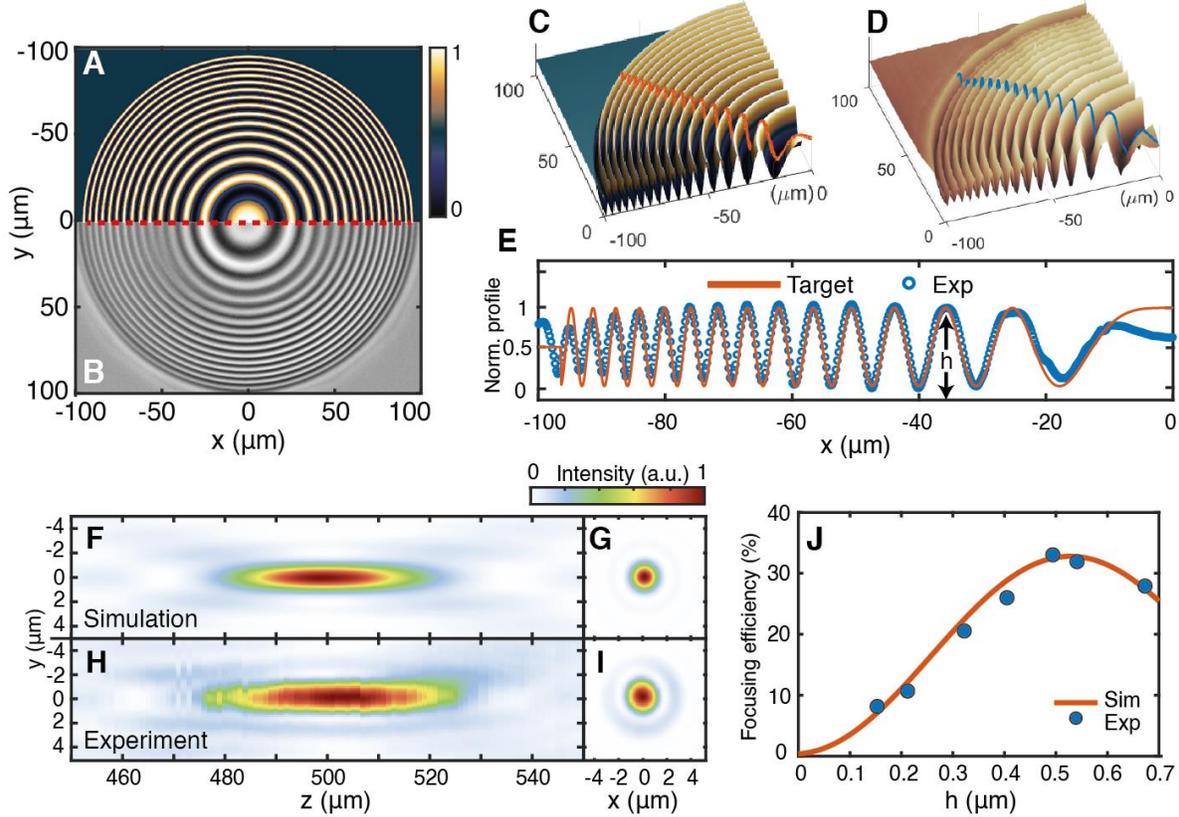

**Fig. 3. Diffractive lens**. (**A**) Target surface (normalized) of a Gabor phase zone plate, designed to focus light of wavelength $\lambda = 633 nm$ at distance $f_0 = 500\ \mu m$ from the surface. (**B**) Scanning Electron Micrographs (SEM) of the holographically structured photoresist surface. (**C** and **D**) Three-dimensional views for a section of the target (C) and AFM measured (D) surface. (**E**) Target and experimental profiles traced along the radial directions shown in (C) and (D), respectively. (**F** to **I**) Comparison of the normalized axial (F) and transverse(G) Point-Spread Functions (PSFs) (simulated for the target lens profile in (A) through Fresnel diffraction integral) and experimental axial (H) and lateral (I) PSFs produced by the experimental surface in (B) under illumination with a collimated beam at design wavelength. (**J**) Comparison of simulated and experimental focusing diffraction efficiency as function of average surface modulation amplitude $h$ (defined in (E)).

Figure 4 shows the possibility of re-shaping our diffractive lens in real-time in the aligned optical system. Figure 4A-C show three different lenses realized one after the other in the same photoresist area. The respective focal field distributions are reported in Fig. 4D-F: as the focal length increases, the numerical aperture decreases and the focused light spot increases. Notably, about 70% shift of the focal position is achieved.

By means of our approach, an imaging system can be realized, able to dynamically provide different magnifications of extended scenes (Fig. 4K-P). Standard zooming systems require the axial movement of at least two lenses to produce a magnified image in the camera plane. In our system instead one of the mechanical motions can be replaced by the lens re-configuration into the needed optical element. This is evident in the magnified images of our Institutions logos. In this case, one lens of the optical system is physically shifted to re-gain the focal position while changing on-demand the focal length of the reconfigurable diffractive lens. The magnification factors reproduce the focal lengths ratios: $f_1/f_0 =1.2$ in Fig. 4I and J; $f_2/f_0=1.6$ for Fig. 4K and L; $f_3/f_0 =2$ for Fig. 4M and N.

An important aspect to highlight for the realized devices is that they are all just made of a structured surface, thinner than the illuminating light wavelength, on a polymer film spin coated on a glass coverslip. The photolithography process to obtain such lightweight and planar devices is fully scalable and compatible with curved substrates, opening new possibilities in functionalize surfaces of objects as diverse as wearable items and vehicle parts.

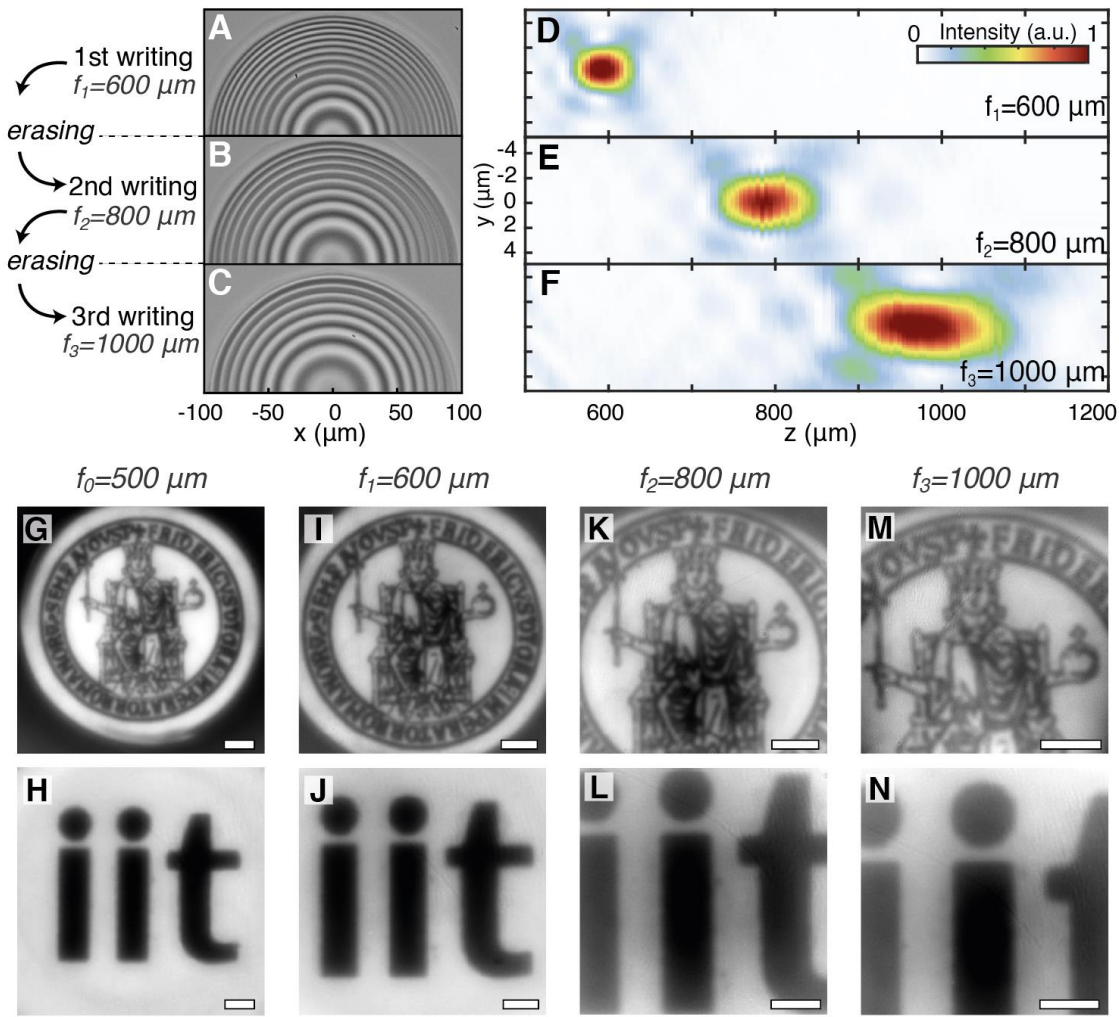

**Fig. 4. Reconfigurable diffractive lenses in the light-path of dynamical imaging systems**. (**A** to **C**) SEM micrographs of a fixed photoresist surface area, holographically re-structured as a diffractive lens of three different focal lengths $f_i$. (**D** to **F**) axial PSFs measured at each reconfiguration step for lenses in (A to C). The focal position axially translates according to the reconfigured surface geometry. (**G** to **N**) Optical images of transparencies depicting our Institution logos produced at increasing magnifications (1.0$X$, 1.2$X$, 1.67$X$ and 2.0$X$), upon the in-place reconfiguration of the diffractive lens focal length. Scale bars are 2.0 $mm$ and 5.0 $mm$ in top and bottom panels, respectively.

## Conclusions

In conclusion, we have proved that diffraction optical elements with efficiency equal to the theoretical efficiency can be realized by direct structuring of the surface of a photosensitive polymer, avoiding any further development step. The realized gratings and lenses can be reshaped completely while aligned in the optical setup. Grating periodicity can be changed; lenses focal length can be tuned; one optical element can be morphed into another optical element with completely different optical functionality, without affecting the alignment of the specific optical setup. More than 100 years after Michelson was reporting about optical effects from imperfect gratings, we show that it is possible to realize optical elements with theory-matching efficiency and practical use, reconfigurable on demand right where and when needed.

# References

1. M. Born, E. Wolf, *Principles of Optics: Electromagnetic Theory of Propagation, Interference and Diffraction of Light* (Cambridge University Press, 1997).

2. J. W. Goodman, *Introduction to Fourier Optics* (Roberts and Company Publishers, 2005).

3. S. M. Kamali, E. Arbabi, A. Arbabi, Y. Horie, M. Faraji-Dana, A. Faraon, Angle-Multiplexed Metasurfaces: Encoding Independent Wavefronts in a Single Metasurface under Different Illumination Angles. *Phys. Rev. X.* **7**, 041056 (2017).

4. D. C. O'Shea, T. J. Suleski, A. D. Kathman, D. W. Prather, *Diffractive Optics* (SPIE, 1000 20th Street, Bellingham, WA 98227-0010 USA, 2003; http://link.aip.org/link/doi/10.1117/3.527861).

5. D. Gabor, A New Microscopic Principle. *Nature.* **161**, 777–778 (1948).

6. S. Banerji, M. Meem, A. Majumder, F. G. Vasquez, B. Sensale-Rodriguez, R. Menon, Imaging with flat optics: metalenses or diffractive lenses? *Optica.* **6**, 805 (2019).

7. K. Huang, F. Qin, H. Liu, H. Ye, C.-W. Qiu, M. Hong, B. Luk'yanchuk, J. Teng, Planar Diffractive Lenses: Fundamentals, Functionalities, and Applications. *Advanced Materials.* **30**, 1704556 (2018).

8. Z. Yang, T. Albrow-Owen, W. Cai, T. Hasan, Miniaturization of optical spectrometers. *Science.* **371** (2021), doi:10.1126/science.abe0722.

9. H. Rubinsztein-Dunlop, A. Forbes, M. V. Berry, M. R. Dennis, D. L. Andrews, M. Mansuripur, C. Denz, C. Alpmann, P. Banzer, T. Bauer, E. Karimi, L. Marrucci, M. Padgett, M. Ritsch-Marte, N. M. Litchinitser, N. P. Bigelow, C. Rosales-Guzmán, A. Belmonte, J. P. Torres, T. W. Neely, M. Baker, R. Gordon, A. B. Stilgoe, J. Romero, A. G. White, R. Fickler, A. E. Willner, G. Xie, B. McMorran, A. M. Weiner, Roadmap on structured light. *J. Opt.* **19**, 013001 (2016).

10. F. Qin, B. Liu, L. Zhu, J. Lei, W. Fang, D. Hu, Y. Zhu, W. Ma, B. Wang, T. Shi, Y. Cao, B. Guan, C. Qiu, Y. Lu, X. Li, π-phase modulated monolayer supercritical lens. *Nature Communications.* **12**, 32 (2021).

11. N. Yu, F. Capasso, Flat optics with designer metasurfaces. *Nat Mater.* **13**, 139–150 (2014).

12. M. Khorasaninejad, A. Ambrosio, P. Kanhaiya, F. Capasso, Broadband and chiral binary dielectric meta-holograms. *Science Advances.* **2**, e1501258 (2016).

13. R. C. Devlin, A. Ambrosio, N. A. Rubin, J. P. B. Mueller, F. Capasso, Arbitrary spin-to-orbital angular momentum conversion of light. *Science*, eaao5392 (2017).

14. A. M. Shaltout, V. M. Shalaev, M. L. Brongersma, Spatiotemporal light control with active metasurfaces. *Science.* **364** (2019), doi:10.1126/science.aat3100.

29. N. Lassaline, R. Brechbühler, S. J. W. Vonk, K. Ridderbeek, M. Spieser, S. Bisig, B. le Feber, F. T. Rabouw, D. J. Norris, Optical Fourier surfaces. *Nature*. **582**, 506–510 (2020).

30. O. Hignette, J. Santamaria, J. Bescos, White light diffraction patterns of amplitude and phase zone plates. *J. Opt.* **10**, 231–238 (1979).

31. S. L. Oscurato, F. Borbone, P. Maddalena, A. Ambrosio, Light-Driven Wettability Tailoring of Azopolymer Surfaces with Reconfigured Three-Dimensional Posts. *ACS Applied Materials & Interfaces*. **9**, 30133–30142 (2017).

32. M. Pasienski, B. DeMarco, A high-accuracy algorithm for designing arbitrary holographic atom traps. *Opt. Express, OE*. **16**, 2176–2190 (2008).

33. R. W. Gerchberg, W. O. Saxton, A practical algorithm for the determination of the phase from image and diffraction plane pictures. *Optik*. **35**, 237–246 (1972).

34. I. Andries, T. Galstian, A. Chirita, Approximate analysis of the diffraction efficiency of transmission phase holographic gratings with smooth non-sinusoidal relief. *Journal of Optoelectronics and Advanced Materials*. **18**, 56–64 (2016).
**Acknowledgments:**

**Funding:** This work has been financially supported by the European Research Council (ERC) under the European Union's Horizon 2020 research and innovation programme "METAmorphoses", grant agreement no. 817794.

**Author contributions:**

Conceptualization: SLO, AA; Methodology: SLO, FR, PM, AA; Investigation: SLO, FR, MS, FB; Supervision: PM, AA; Writing – original draft: SLO, AA; Visualization: SLO, FR, MS, FB, PM, AA.

**Competing interests:** Authors declare no competing interests.
**Supplementary Materials:**

Materials and Methods

Figures S1-S8

Tables S1

References (*31-34*)

# Supplementary Materials for

# Shapeshifting diffractive optical devices

S.L. Oscurato, F. Reda, M. Salvatore, F. Borbone, P. Maddalena, A. Ambrosio

Correspondence to: stefanoluigi.oscurato@unina.it; antonio.ambrosio@iit.it;

## Materials and Methods

Photoresist synthesis and characterization

The photoresist used in this work is an azobenzene-containing polymer (azopolymer) in amorphous state. All reagents were purchased from Merck and used without further purification. The azopolymer was synthesized, purified and characterized as previously reported (Mw = 27000; phase sequence: Glass 67 °C Nematic 113 °C Isotropic; $\lambda_{max}$ = 350 nm) (*17*, *31*). The solution for film deposition was prepared by dissolving 70 mg of the polymer in 0.50 ml of 1,1,2,2-tetrachloroethane and filtered on 0.2 µm PTFE membrane filters. The desired film thickness (typically $1.0 \pm 0.1\ \mu m$) was obtained by spin coating the solution on 24x60 mm cover slides at 300 rpm for 4 minutes. In the final stage, the samples were kept under vacuum at room temperature for 24 h to remove solvent traces.
Refractive index of the fabricated film was measured via ellipsometry. Measured values at some relevant wavelengths (633 nm, 532 nm, 488 nm) are: $n_{633} = 1.70$; $n_{532} = 1.74$; $n_{488} = 1.78$.

Holographic illumination setup

The experimental configuration for the azopolymer surface relief inscription is based on a phase-only Computer-Generated Holograms (CGHs) scheme. Its schematic representation is shown in Fig. S1. A laser diode source (Cobolt Calypso) emits a TEM$_{00}$ beam at wavelength $\lambda = 491\ nm$ and, after a beam expander, is phase-modulated by a computer-controlled reflective phase-only Spatial Light Modulator (SLM, Holoeye Pluto). The modulated beam is propagated through a $4f$ lenses system with the input plane located in the SLM plane. The output plane coincides with the back focal plane of an infinity-corrected long-working distance 50X objective (Mitutoyo), with numerical aperture $NA = 0.55$. This configuration allows the reconstruction of a structured intensity pattern in the focal plane of the objective (where the photoresist is placed).

Arbitrary intensity patterns can be generated imposing the proper phase profile (*kinoform*) for the beam in the SLM plane. The phase hologram is calculated according to the Fourier transform relations (*2*) existing between $4f$ system conjugate planes. The focal lengths of the lenses L$_3$ and L4 (Fig. S1) are chosen in order to maximize the spatial resolution in the hologram reconstruction planes (*24*). This choice also defines the diameter (~200 $\mu m$) of the accessible circular area in the objective front focal plane, which can be used to structure the photoresist surface in a single illumination step. The position of the sample near the objective focal region is accurately controlled by means of a x-y-z translation stage. Average intensity in the range $12.7 - 14.0\ W/cm^2$ and circular polarization are used for the structuration of the azopolymer surface. For visual inspection, and proper focusing of the holographic pattern on the photoresist surface, a beam splitter placed in the light-path redirects the light retroreflected by the surface and re-collimated through the objective toward a tube lens. This lens forms an image of the holographic pattern in the second focal plane, where a CCD camera is positioned to observe the surface plane in real-time (Fig. S1).

When needed, an additional diode laser beam at $405\ nm$ illuminates the photoresist film from the substrate (glass microscope coverslip) side. The beam has circular polarization and different intensity levels depending on its intended function. When the intensity is in the range $0.4 - 0.8\ W/cm^2$, the beam favours the surface structuring process, acting as a writing *assisting beam.* At intensity higher than $0.9\ W/cm^2$, its absorption causes the erasure of previously inscribed surface structures, acting as an *erasing beam* (see also Fig. S4).

Algorithm for calculation of phase-only holograms

The Iterative Fourier Transform Algorithm (IFTA) used for the calculation of the phase profile of the writing beam in the SLM plane is the Mixed Region Amplitude Freedom (MRAF) algorithm (*32*), based on an extension of the standard Gerchberg-Saxton (GS) algorithm (*33*). The IFTA have been implemented in MATLAB, using Fast Fourier Transform (FFT). In the first step of the calculation, a target 8-bit grayscale image (see also Fig. S2), representing the desired intensity pattern (e.g. sinusoidal) to be reconstructed in the photoresist plane, is analytically defined. Next, with an iterative loop based on direct and inverse FFTs, the algorithm returns a grayscale (8 bit, 256 phase levels) image (1080 X 1920 px) for the phase profile (the *kinoform*;) to be imposed on the writing beam in the SLM plane.

Optical real-time characterization of structured surfaces

For real-time observation of the photoresist surfaces during the structuration process, the holographic setup was integrated with a collimated white LED source for a bright-field transmission microscopy system. Scattered light from the sample surface is collected by the objective. Real time image of the sample surface is obtained with the same configuration for holographic pattern acquisition of Fig. S1. The retroreflected holographic pattern, propagating along the same light path, can be eventually discarded in the imaging using a low pass filter before the camera.

For the collection of real-time diffraction patterns shown in Fig. 1, a second beam splitter redirects part of the white light emerging from the tube lens through a positive lens realizing a $2f$ configuration with the tube lens. A second CCD camera, positioned in the second focal plane of the lens, provides a Fourier transform image of the structured surface, corresponding to the far-field diffraction pattern.

A second configuration allows to produce real-time diffraction patterns from photoresist structured surface using a He-Ne as probe beam, whose wavelength ($\lambda = 633$ nm) lies outside the absorption band of the azopolymer (*17*, *31*), not interfering then with the writing process. The probe beam is focused in the back focal plane of the objective using the beam splitter and the tube lens. The collimated beam, emerging from the objective, illuminates the same photoresist area simultaneously structured by the holographic writing beam. For the time-dependent analysis of diffraction efficiencies produced by one-dimensional diffraction gratings (Fig. 1E of the main text), photodiodes are used to measure the light power of the probe beam diffracted in far-field in the first diffraction order.

Morphological characterization of structured surfaces

Topographic characterization of photoresist surfaces is performed using Atomic Force Microscopy (AFM) and Scanning Electron Microscopy (SEM).
For AFM measurements, a WITec Alpha RS300 microscope is used. The AFM is operated in tapping mode using a cantilever with $75\ kHz$ resonance frequency. The maximum scanned area has a size of $100\ X\ 100\ \mu m^2$.
Scanning Electron Microscopy images are acquired with a field-emission gun (FEG–SEM) FEI/ThermoFisher Nova NanoSEM 450 microscope. Samples are sputtered with a layer of Au/Pd using a Denton Vacuum Desk V TSC coating system prior to observation.

Focusing efficiency and axial Point Spread Function measurements

The measurement of focusing efficiency reported in Fig. 3J of the main text was performed using a He-Ne laser beam, incident on the diffractive lens from the substrate side. Light focused by the lens is collected though the same microscope imaging system used for brightfield surface observation. An iris and a power meter are used to intercept and measure only the light power transmitted through an aperture of the same diameter as the image of the structured area. Focusing efficiency is calculated as the ratio of the light power measured when the lens is written on the surface with respect to the same measurement with no structures in the photoresist area.
For the measurement of the axial point spread function of the reconfigured diffractive lenses, the iris is removed and the images of the transmitted He-Ne laser beam obtained axially translating the diffractive lens with $1\ \mu m$ (Fig. 3) or $5\ \mu m$ (Fig. 4) steps are collected with the CCD in the focal plane of the tube lens.

# Supplementary Figures

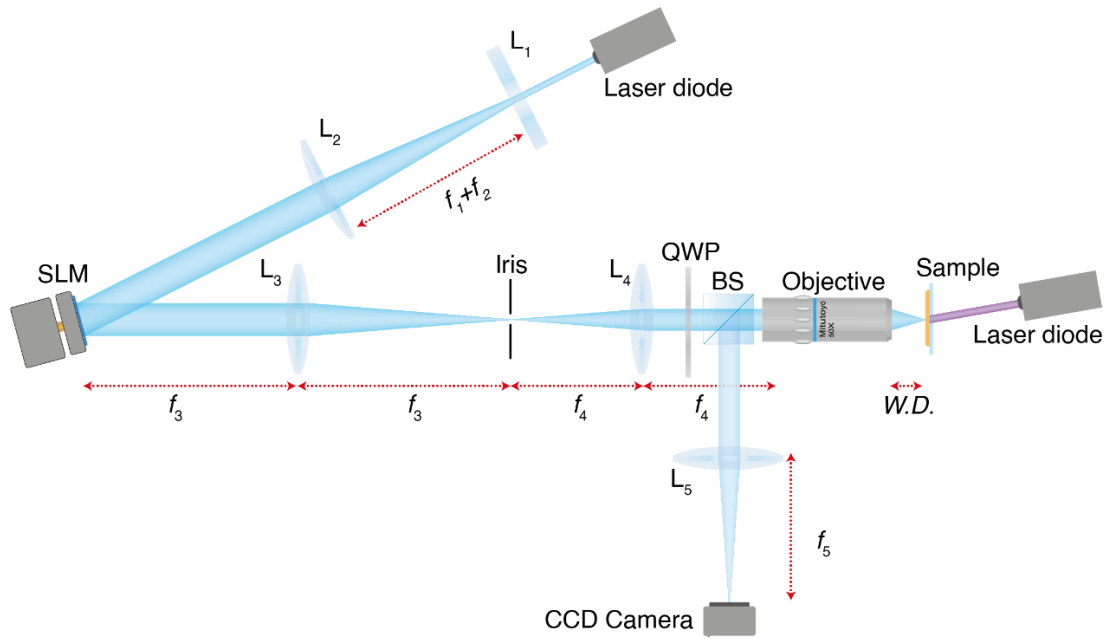

**Fig. S1**
**Setup for holographic structuration of photoresist surface.**
The writing laser beam at wavelength $\lambda = 491\ nm$, after a beam expander (lenses $L_1$ and $L_2$), is modulated by a reflective phase-only Spatial Light Modulator (SLM). The SLM is placed in the first focal plane of a bi-convex lens ($L_3$), with a focal length of $f_3 = 300\ mm$, realizing a $2f$ configuration with the plane of an iris placed in the second focal plane. The iris allows the spatial filtering of the beam, blocking all the undesired lower-intensity diffraction orders and the un-modulated light emerging from the SLM. The laser beam is re-collimated using a bi-convex lens $L_4$ (with focal length $f_4 = 175\ mm$), which projects a rescaled version of the optical field of the SLM plane in its second focal plane. That plane coincides with the back focal plane of an infinity corrected "50X Mitutoyo" objective (OBJ). The structured holographic intensity distribution is then reconstructed and focused on the objective focal plane (objective working distance WD $13\ mm$) where the photoresist film is placed. Writing beam is circular polarized using a quarter wave plate (QWP). A 70/30 beam splitter (BS), placed between objective and $L_4$ lens, allows the collection of part of retroreflected writing light which is collected and recollimated by the objective. Finally, the tube lens ($L_5$) ($f_5 = 200\ mm$) focuses the image of the objective focal plane on a "DCC3240M Thorlabs" camera (CCD). The assisting (or erasing) beam at $\lambda = 405\ nm$ is circularly polarized before being redirected in the writing\erasing area at near-normal incidence.

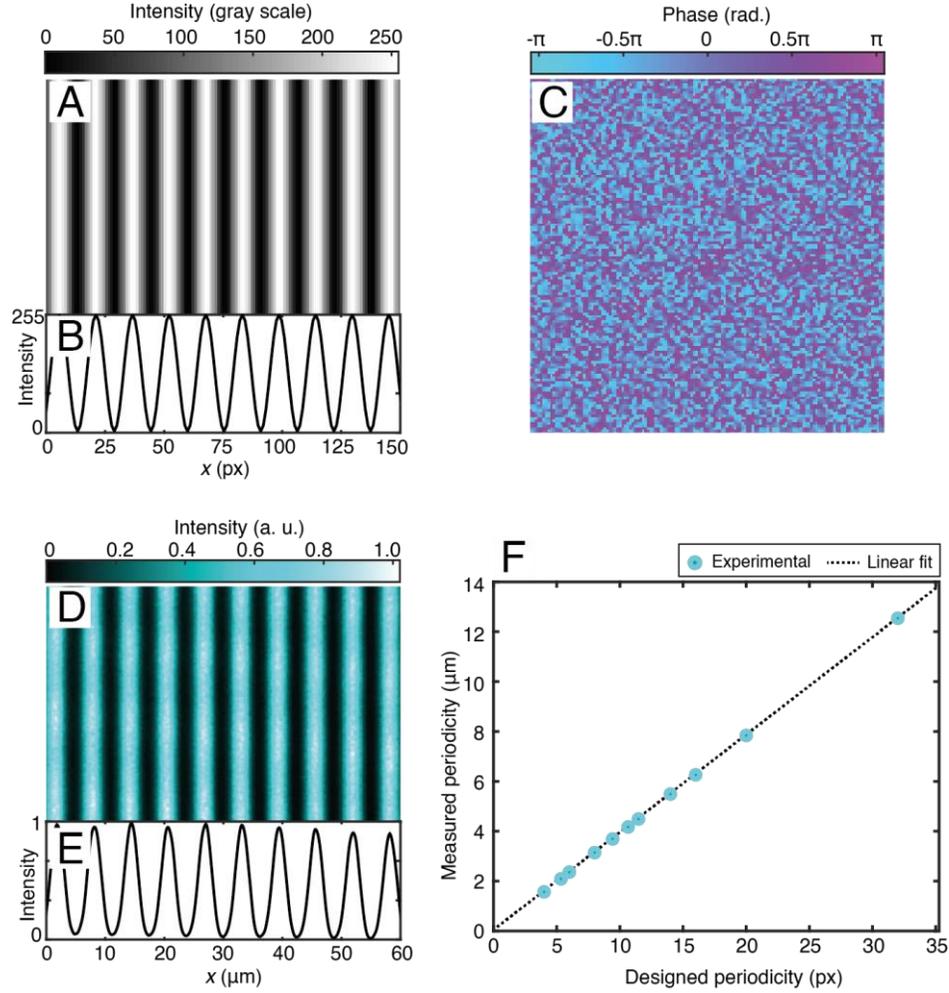

**Fig. S2**

**Design of holographic sinusoidal intensity patterns**

The fundamental aspects of the holographic design and calibration for sinusoidal patterns in our system are described. (**A**) Typical grayscale 8-bit target image used for the calculation of a one-dimensional holographic sinusoidal intensity patterns. (**B**) Target sinusoidal profile. (**C**) The *Mixed Region Amplitude Freedom* (MRAF) algorithm calculates the phase profile (the kinoform) for the writing laser beam in the Spatial-Light Modulator plane necessary to reconstruct the desired intensity distribution in photoresist plane. Kinoforms are in the range $[-\pi, \pi]$ interval and are encoded in 8-bit grayscale images, directly transferred to the modulator. (**D**) Reconstructed intensity pattern in the photoresist plane: image is acquired using tube lens-CCD camera system, collecting retroreflected light from the photoresist plane with a CCD. The image is the result of averaging the holographic speckle noise over 1000 independent images of the pattern. CCD counts are renormalized in $[0, 1]$ interval. (**E**) Reconstructed intensity pattern profile. (**F**) Measured periodicity of different reconstructed sinusoidal intensity patterns as function of the designed pitch in the target image. The slope of the line trend defines the calibration of physical dimensions of patterns in the photoresist plane with respect to the analytically designed target images.

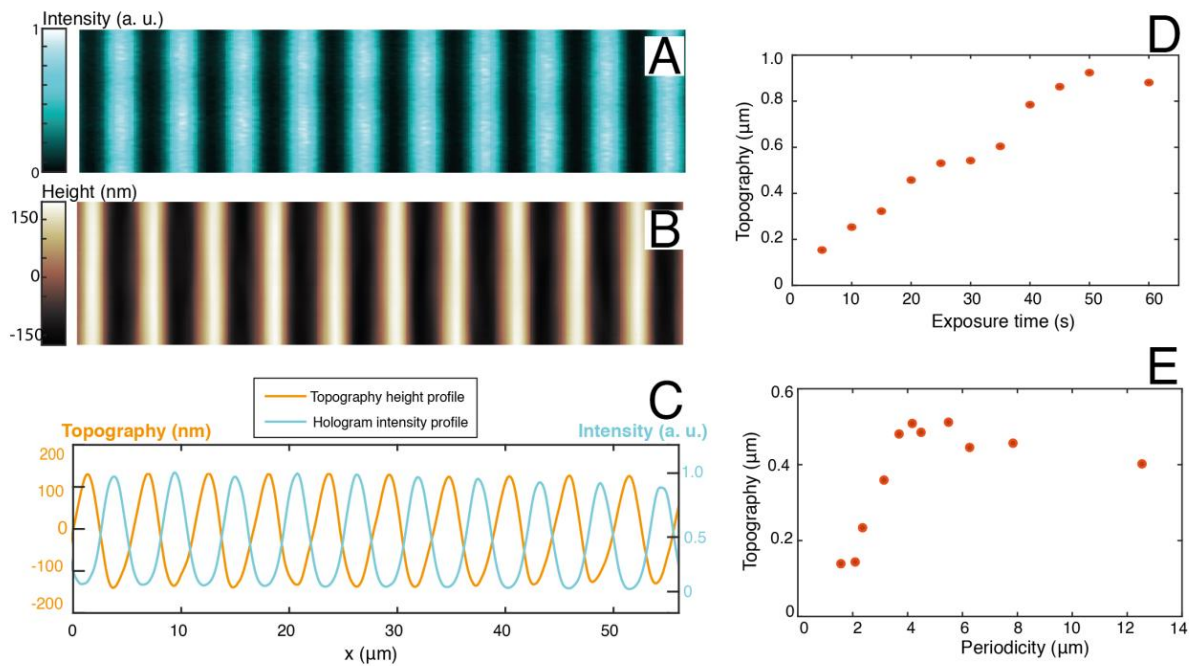

**Fig. S3**

**Sinusoidal illumination patterns and azopolymer photo-structuration**

(**A**) Sinusoidal intensity pattern, of periodicity 5.49 $\mu m$, used as example of illumination pattern for the direct inscription of sinusoidal gratings on the photoresist surface. (**B**) AFM micrograph of the structured photoresist surface. (**C**) Comparison between measured AFM profile and the holographic intensity pattern profile. (**D**) Measured relief amplitude obtained for different exposure times at fixed illumination intensity. The linear trend shows a direct proportionality between relief height and exposure time. (**E**) Dependence of the relief amplitude with respect to the periodicity of the sinusoidal illumination pattern. The relief inscription efficiency can be assumed independent of the sinusoidal pitch for relatively hight values. Smaller values determine a loss of efficiency due to limitation in holographic technique related to both spatial resolution limits in holographic setup and the discretized description of target images. For large periodicities, the dependence on light intensity gradient strength in equation S9 causes a reduction of inscription efficiency. If needed, these effects could be eventually compensated by tuning the exposure time for each sinusoidal periodicity.

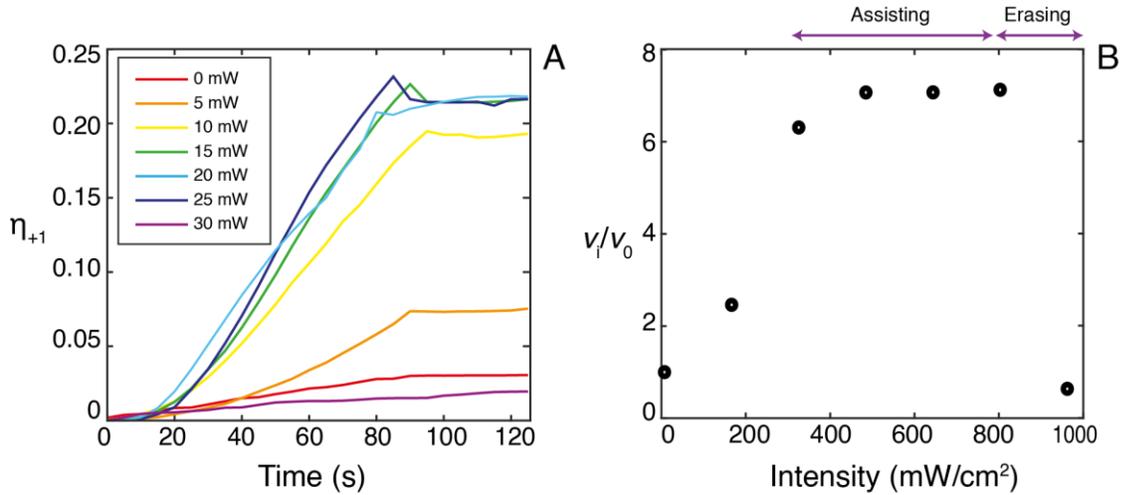

**Fig. S4**
**Influence of assisting/erasing beam on relief inscription dynamics**

Influence of the assisting beam on the sinusoidal structuration process, studied in real time exposing the azopolymer free surface to a sinusoidal illumination pattern. The structuration is monitored using real time diffraction efficiency measurement. (**A**) Real-time diffraction curves of the $+1$ diffraction order, recorded for different power of the assisting beam (at $405\ nm$) incident (collimated, diameter $\sim 2.0\ mm$) on the photoresist during the holographical writing. The exposure time is fixed at $90\ s$. According to the diffraction efficiency dynamics (A) and velocity (B) (extracted from the slope of the linear portion of the trends in A), surface relief inscription efficiency increases under the influence of low intensity (in the range $400 - 800\ mW/cm^2$) *assisting* beam. At higher intensity ($> 900\ mW/cm^2$), surface structuration process is slowed down by the $405\ nm$ beam, until reaching the situation where the intensity is so high to completely avoid the formation of surface structures or erase previously inscribed ones (*erasing* configuration).

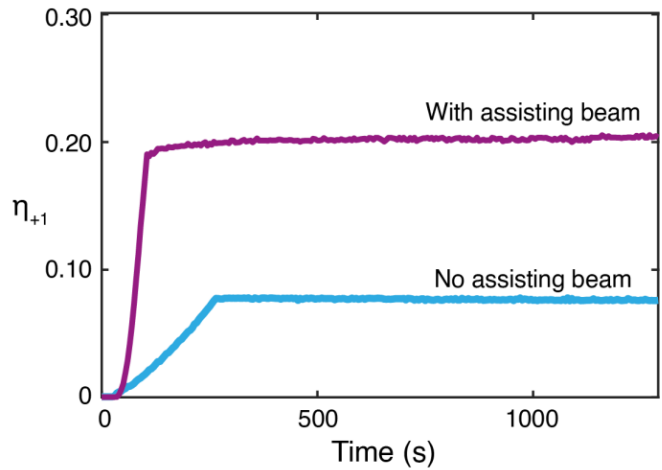

**Fig. S5**
**Surface relief stability over time**

The stability of written structures on the photoresist surface are analyzed through the measurement of real-time first order diffraction efficiency of a sinusoidal diffraction grating. After the initial efficiency rising under illumination, the writing beam (and the assisting beam) is switched off, while the monitoring of diffraction of the probe beam is continued for 30 minutes. Both the structures inscribed with and without the assisting beam, result stable over time, indicating a diffraction behavior entirely determined by the surface relief pattern (e.g. no contribution from eventual birefringence gratings originated by possible photo-reorientation of photoresist chromophores). The stability of the patterns is preserved for years by maintaining the samples at normal room conditions.

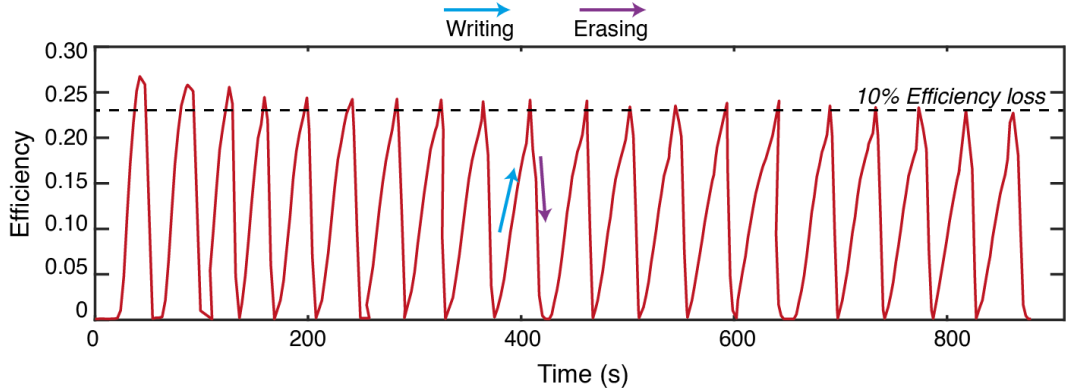

**Fig. S6**
**Analysis of multiple photoresist writing and erasing cycles**

To characterize the repeatability surface remodulation process, the photoresist surface is cyclically structured and erased with a sinusoidal intensity pattern (with periodicity $\Lambda = 5.5\ \mu m$ and assisting beam). During each writing and erasing cycle, the dynamical time-dependent efficiency of the He-Ne probe beam diffracted in +1 order is measured with a photodiode. In the experiment, at least 15 writing and erasing cycles (with approximately equal writing and erasing time) are obtained without significant writing and erasing efficiency reduction (only a loss of 10% in the maximum diffraction efficiency value is measured between the first and last writing step).

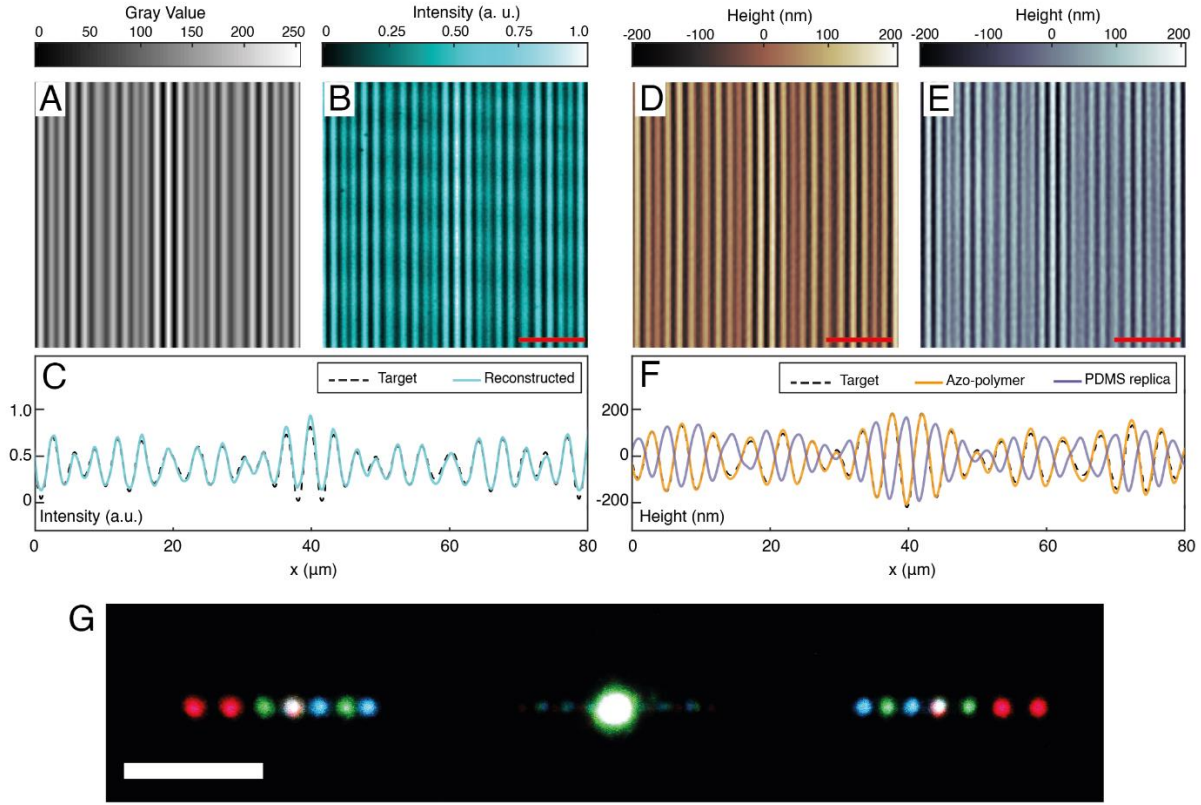

**Fig. S7**
**Design and realization of the RGB grating**

RGB grating design, encoding in the holographic pattern and lithographic process on the photoresist with subsequent pattern transfer on a Polydimethylsiloxane (PDMS) mold. **(A)** Target image designed as superposition of three sinusoidal patterns with equal amplitudes and spatial periodicities $\Lambda_1 = 3.13\ \mu m$, $\Lambda_2 = 3.42\ \mu m$, $\Lambda_3 = 4.05\ \mu m$, designed to diffract three different light wavelengths ($\lambda_1 = 633\ nm$, $\lambda_2 = 532\ nm$ and $\lambda_3 = 488\ nm$) under the same angle in the first diffraction order. Target image is encoded in a grayscale 8-bit digital image. **(B)** Reconstructed intensity pattern in the photoresit plane obtained with the Computer-Generated Hologram systems, using the gray scale image in (A) as input data of the kinoform calculation process of the MRAF algorithm. The image in (B) is the result of averaging the holographic speckle noise over 1000 independent images of the pattern collected by the CCD (CCD counts are mapped in $[0, 1]$ interval). **(C)** Comparison between the lateral profiles of the target image (normalized in $[0, 1]$ interval) and the experimental holographic intensity pattern. **(D)** AFM scan of the photoresist surface (from Fig. 2B in the main text) after being exposed to the averaged (by refreshing the kinoform on the SLM at 20 Hz) holographic intensity pattern. **(E)** AFM micrograph of the PDMS replica obtained with standard molding process, using the photoresist surface in (D) as master. The PDMS (Sylgard 184, Dow Corning) mixture was prepared by mixing the precursor and the curing agent in a 10:1 weight ratio. After degassing in a vacuum chamber, the PDMS mixture was gently poured onto the azopolymer film without further surface treatment and cured at $30\ °C$ for $6\ hours$ before being carefully released from the film. **(F)** Comparison between topographic profiles of the photoresist surface and PDMS mold (inverse replica) with respect to the target designed pattern: the graph shows an optimum agreement between the surface topography of the photoresist and the target. The total height amplitude is preserved in the pattern transfer (with a measured loss of less than 8%). The fabricated PDMS is fully compatible with standard lithographic replica molding processes for the transfer of the surface pattern on other materials. Scale bars in A-B-D-E are $20~\mu m$. **(G)** Full view of the RGB diffraction pattern of Fig. 2D of the main text, produced by the surface in (**D**) from simultaneous illumination of the DOE with three collinear laser beams at design wavelengths ($\lambda_1 = 633~nm; \lambda_2 = 532~nm, \lambda_3 = 488~nm$). Scale bar in (G) is $0.5~cm$.

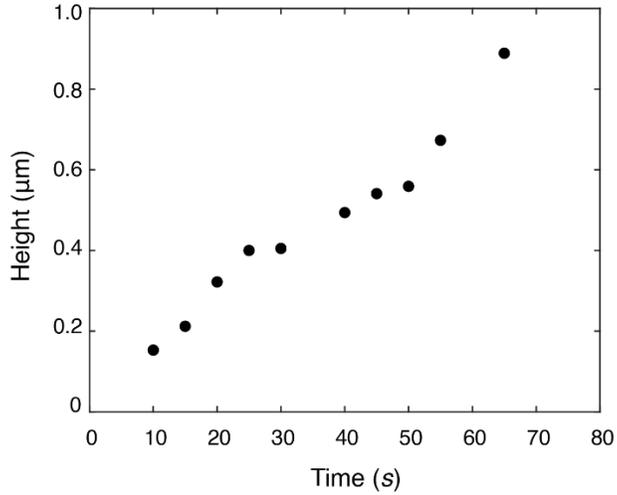

**Fig. S8**
**Tuning of Gabor phase zone plate relief amplitude**

Graph shows the measured average relief amplitude $h$ in the radial surface profile of the experimental Gabor phase zone plate lens reported in Fig. 3 of the main text measured for different exposure times. A linear amplitude increase is obtained for increasing exposure time. The target amplitude $h \approx 530 \; nm$ (producing the theoretical maximum focusing efficiency) is realized in 50s of exposure at the holographic pattern in the presence of the assisting beam.

| RGB grating $N = 3$ | Sinusoidal pitch $\Lambda_i$ (µm) | (3.1 – 3.4 – 4.1) |
|---|---|---|
| | Relative amplitude $A_i$ | 1 |
| | Relative phase $\varphi_i$ (rad) | (0 – 0 – 0) |
| | In plane orientation $\gamma_i$ (deg) | (0 – 0 – 0) |
| | $T$ (s) | 60 |

| Quasi-crystal $N = 6$ | Sinusoidal pitch $\Lambda_i$ (µm) | (3.1 – 3.1 – 3.1 – 3.1 – 3.1 – 3.1) |
|---|---|---|
| | Relative amplitude $A_i$ | 1 |
| | Relative phase $\varphi_i$ (rad) | (0 – 0 – 0 – 0 – 0 – 0) |
| | In plane orientation $\gamma_i$ (deg) | (0 – 30 – 60 – 90 – 120 – 150) |
| | $T$ (s) | 120 |

| Spiral quasi-crystal $N = 6$ | Sinusoidal pitch $\Lambda_i$ (µm) | (3.1 – 3.9 – 4.7 – 5.5 – 6.3 – 7.1) |
|---|---|---|
| | Relative amplitude $A_i$ | 1 |
| | Relative phase $\varphi_i$ (rad) | (0 – 0 – 0 – 0 – 0 – 0) |
| | In plane orientation $\gamma_i$ (deg) | (0 – 30 – 60 – 90 – 120 – 150) |
| | $T$ (s) | 120 |

| Blazed grating $N = 6$ | Sinusoidal pitch $\Lambda_i$ (µm) | (12.5 – 6.3 – 4.2 – 3.1 – 2.5 – 2.1) |
|---|---|---|
| | Relative amplitude $A_i$ | $1/i$ |
| | Relative phase $\varphi_i$ (rad) | (0 – $\pi$ – 0 – $\pi$ – 0 – $\pi$) |
| | In plane orientation $\gamma_i$ (deg) | (0 – 0 – 0 – 0 – 0 – 0) |
| | $T$ (s) | 120 |

**Table S1.**
**Design parameters for the realization of the complex diffraction gratings**
Table of design parameters for the realization of the diffractive surfaces shown in Figure 2(B-F-H-J) of the main text, named in the first column of each sub-table. For every surface is reported the number $N$ of combined sinusoidal functions included in the structure design. The remaining rows report the sinusoidal pitch $\Lambda_i$, the relative amplitude $A_i$, the relative phase $\varphi_i$, the in-plane orientation $\gamma_i$ of each of the $N$ components of the designed geometry. Total exposure time $T$ is related with the total relief amplitude $h$.